\documentclass{article}
\usepackage{epsfig}

\begin{document}
\title{Forward Jets and BFKL at Hadron Colliders%
\thanks{Presented at DIS2002}%
}
\author{Jeppe R.~Andersen
}
\hspace*{\fill}\parbox[t]{4cm}{
IPPP/02/38\\
DCPT/02/76\\July 1, 2002}
\begin{center}
\textbf{\Large Forward Jets and BFKL at Hadron Colliders}\footnote{Presented
  at the X International Workshop on Deep Inelastic Scattering (DIS2002).}\\
\vspace{1.cm}

{Jeppe R.~Andersen}\\
\vspace{.2cm}
{\sl Institute for Particle Physics Phenomenology\\
University of Durham\\
Durham, DH1 3LE, U.K.}\\
\end{center}
\begin{abstract}
  We present results on dijet and $W$+dijet production at hadron colliders
  obtained by supplementing the leading log BFKL resummation with energy and
  momentum conservation. For pure dijet production, the inclusion of the BFKL
  radiation in the energy conservation leads to a decrease in the parton flux
  sufficient to counter-act the expected exponential increase in the cross
  section obtained for the partonic cross section. Other BFKL signatures
  such as the dijet azimuthal angle decorrelation do still survive.
\end{abstract}
  
\section{Introduction}
When confronting BFKL with data, it must be remembered that the analytic
leading log (LL) BFKL resummation\cite{Kuraev:fs} makes some approximations
which, even though formally subleading, can be numerically important at
present collider energies. These approximations include: $a)$ The BFKL
resummation is performed at fixed coupling constant.  $b)$ Because of the
strong rapidity ordering any two-parton invariant mass is large. Thus there
are no collinear divergences in the LL resummation in the BFKL ladder; jets
are determined only at tree-level and accordingly have no non-trivial
structure.  $c)$ Finally, energy and longitudinal momentum are not conserved,
since the momentum fraction $x$ of the incoming parton is reconstructed
without the contribution to the total energy from the radiation of the BFKL
ladder.  Therefore, the analytic BFKL approach systematically underestimate
the exact value of the $x$'s, and can thus grossly overestimate the parton
luminosities. In fact, for dijet production (at a hadron collider) with a
BFKL gluon exchange in the $t$-channel we have
\begin{equation}
  x_{a(b)}=\frac{P_{a\perp}}{\sqrt s}e^{(-)y_a}+\frac{P_{b\perp}} {\sqrt
  s}e^{(-)y_b} + 
  \sum_{i=1}^{n}\frac{k_{i\perp}}{\sqrt s}e^{(-)y_i}\label{eq:xs},
\end{equation}
where the minus sign in the exponentials of the right-hand side applies to
the subscript $b$ on the left-hand side. $x_a,x_b$ is the Bjorken $x$ of the
incoming partons, and $(P_{a\perp},y_a),(P_{b\perp},y_b)$ is the transverse
momentum and rapidity of the two leading dijets. The sum is over the number
$n$ of gluons emitted from the BFKL chain, each with transverse momentum
$k_{i\perp}$ and rapidity $y_i$. It is this last contribution to the energy
and longitudinal momentum conservation that is inaccessible in the standard
analytic approach to LL BFKL, since the BFKL equation is solved by summing over
any number of gluons radiated and integrating over the full allowed rapidity
ordered gluon phase space. Considering Mueller-Navelet dijet
production\cite{Mueller:1987ey}, a comparison of three-parton production to
the truncation of the BFKL ladder to ${\cal O}(\alpha_s^3)$ shows that the LL
approximation leads to sizable violations of energy-momentum
conservation~\cite{DelDuca:1995ng}.

We will, in the following, report on studies of the effects of including
energy and momentum conservation in the LL BFKL evolution.

\section{Monte Carlo Approach to Studying the BFKL Chain}
A Monte Carlo approach to studying the BFKL gluon exchange was first reported
in Ref.~\cite{Schmidt:1997fg,Orr:1997im} and the details of the formalism
will not be repeated here. The basic idea of the Monte Carlo BFKL model is to
solve the BFKL equation while maintaining information on each radiated gluon.
This is done by unfolding the integration over the rapidity ordered BFKL
gluon phase space by introducing a resolution scale $\mu$ discriminating
between resolved and unresolved radiation. The latter combines with virtual
corrections to form an IR safe integral. Thereby the solution to the BFKL
equation is recast in terms of phase space integrals for resolved gluon
emissions, with form factors representing the net effect of unresolved and
virtual emissions. Besides being necessary for calculating the impact on the
parton flux by including energy and momentum conservation, this approach also
allows for further studies of the details of the BFKL radiation, and for the
effects of the running of the coupling to be added to the LL evolution.

\section{BFKL Signatures in Dijet Production}
\begin{figure}[htbp]
  \centering
  \epsfig{width=8cm,file=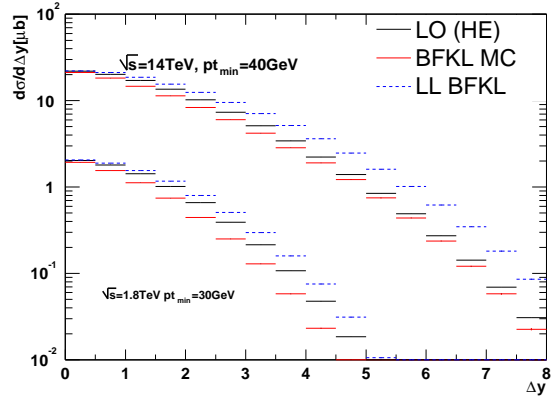}
  \caption{Mueller-Navelet dijet cross sections calculated for the
    high-energy limit of leading order QCD and for LL BFKL, both in the
    standard LL approach and this supplemented with energy-momentum
    conservation (BFKL MC).}
  \label{fig:dijetxsec}
\end{figure}

\begin{figure}[htbp]
  \centering
  \epsfig{width=8cm,file=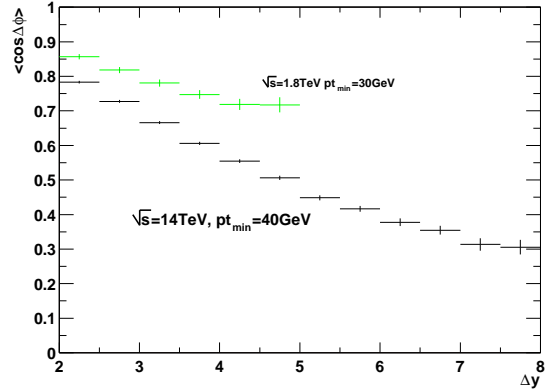}
  \caption{Dijet angular decorrelation of Mueller-Navelet dijets calculated
    for energy-momentum conserving LL BFKL. The leveling out of the
    decorrelation at higher values of the rapidity separation is a result of
    the available phase space restricting further radiation from the BFKL
    chain.}
  \label{fig:dijetdecor}
\end{figure}

The main result of the study \cite{Andersen:2001kt} is that the contribution
of the BFKL gluon radiation to the parton momentum fractions (at LHC
energies) lowers the parton flux in such a way as to approximately cancel the
rise in the subprocess cross section with increasing dijet rapidity
separation ($\hat\sigma_{jj} \sim \exp(\lambda\Delta y$)) predicted from the
standard BFKL approach (see Figure~\ref{fig:dijetxsec}). This strong pdf
suppression is due to the dijet production being driven by the gluon pdf,
which is very steeply falling in $x$ for the region in $x$ of interest. This
means that even the slightest change in $x$ has a dramatic impact on the
parton flux.
The leading-order QCD prediction for the hadronic dijet cross section is
therefore only slightly modified when including BFKL evolution of the
$t$-channel gluon to an almost no-change situation. However, other BFKL
signatures such as the dijet azimuthal angle decorrelation do still survive
(see Figure~\ref{fig:dijetdecor}).

\section{BFKL Signatures in $W+2$jet Production}
\begin{figure}[htbp]
  \centering
  \epsfig{width=8cm,file=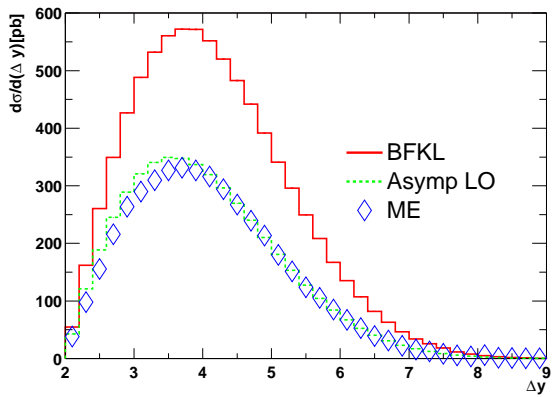}
  \caption{The $W+2$-jet production rate as a function of the
    rapidity interval between the jets $\Delta y$ with the following cuts
    $y_W,y_{j_2}\ge1, y_{j_1}\le -1$ or $y_W,y_{j_2}\le-1, y_{j_1}\ge 1$. The
    diamonds are the leading order production rate; the dashed curve is the
    production rate in the high-energy limit; the solid curve includes the
    BFKL corrections taking energy/momentum conservation into account.}
  \label{fig:w2jet}
\end{figure}

Although at hadron colliders the simplest process for studying BFKL effects
is the production of dijets with large rapidity separation, the formalism
also applies to the production of more complicated forward final states. One
of the forward Mueller-Navelet jets can be replaced by a $W$--jet pair, which
also provides a testing ground for BFKL signatures~\cite{Andersen:2001ja}.
In fact, the suppressing effect of the BFKL gluon radiation on the pdfs is
less pronounced in this case, since requiring a $W$ in the final state at
means (at leading order) that at least one of the initial state partons must
be a {\it quark}, with a less steeply falling pdf. This means that the BFKL
rise in the partonic cross section is not compensated to the same extent as
in the dijet case. In fact, we find that in this case the cross section for
the process including a BFKL gluon exchange is higher than the leading order
cross section, thanks to the relative flatness of the quark pdf in the
relevant region in $x$ (see Figure~\ref{fig:w2jet}).

\begin{figure}[htbp]
  \centering
  \epsfig{width=8cm,file=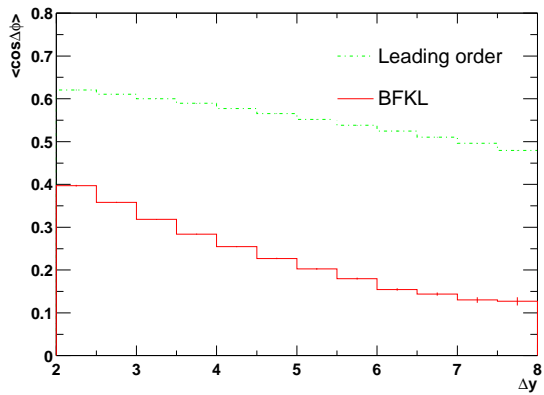}

  \caption{The average azimuthal angle between the two jets in $W$+2jet
    production as a function of the rapidity interval between them. Same cuts
    applied as in Fig.~\ref{fig:w2jet}.}
  \label{fig:w2jetangle}
\end{figure}

In the case of $W$+2jet production, there will be some decorrelation in
azimuthal angle between the two jets already at leading order because of the
radiation of the $W$. However, a BFKL gluon exchange will increase this
decorrelation\cite{Andersen:2001ja} significantly (see
Figure~\ref{fig:w2jetangle}).

\section{Acknowledgments}
The author would like to thank V.~Del Duca, F.~Maltoni, S.~Frixione,
C.~Schmidt, and W.J.~Stirling for a fruitful collaboration and many
stimulating discussions.

\end{document}